\documentclass[journal]{IEEEtranTICPS}
\usepackage{cite}
\usepackage{picinpar}
\usepackage{graphicx}
\usepackage{url}
\usepackage{flushend}
\usepackage[latin1]{inputenc}
\usepackage{colortbl}
\usepackage{soul}
\usepackage{multirow}
\usepackage{pifont}
\usepackage{color}
\usepackage{alltt}
\usepackage[hidelinks]{hyperref}
\usepackage{enumerate}
\usepackage{siunitx}
\usepackage{breakurl}
\usepackage{epstopdf}
\usepackage{pbox}
\usepackage{amsmath,amssymb,amsfonts}
\usepackage{amsthm}
\usepackage{calligra}
\usepackage{algorithmicx}
\usepackage[linesnumbered,ruled,vlined]{algorithm2e}
\usepackage{bm}
\usepackage{epstopdf}
\usepackage{caption}
\usepackage{comment}
\usepackage{booktabs}
\usepackage{subfigure}
\usepackage{breqn}
\usepackage{filecontents} 
\usepackage{flushend}

\newtheorem{definition}{Definition}
\newtheorem{remark}{Remark}

\begin{document}
\title{Sustainable Diffusion-based Incentive Mechanism for Generative AI-driven Digital Twins in Industrial Cyber-Physical Systems}

\author{
	\vskip 1em
	
	Jinbo Wen, Jiawen Kang, Dusit Niyato, \textit{Fellow, IEEE}, Yang Zhang, and Shiwen Mao, \textit{Fellow, IEEE}

	\thanks{
    This research was supported by the National Natural Science Foundation of China (NSFC) under Grants No. 62102099, No. U22A2054, and No. 62071343, Guangzhou Basic Research Program under Grant 2023A04J1699, Guangdong Basic and Applied Basic Research Foundation under Grant 2023A1515140137, and Collaborative Innovation Center of Novel Software Technology and Industrialization. This research was also supported by the National Research Foundation, Singapore, and Infocomm Media Development Authority under its Future Communications Research \& Development Programme, Defence Science Organisation (DSO) National Laboratories under the AI Singapore Programme (FCP-NTU-RG-2022-010 and FCP-ASTAR-TG-2022-003), Singapore Ministry of Education (MOE) Tier 1 (RG87/22), the NTU Centre for Computational Technologies in Finance (NTU-CCTF), Seitee Pte Ltd, and the RIE2025 Industry Alignment Fund - Industry Collaboration Projects (IAF-ICP) (Award I2301E0026), administered by A*STAR, as well as supported by Alibaba Group and NTU Singapore through Alibaba-NTU Global e-Sustainability CorpLab (ANGEL).
    
	J. Wen and Y. Zhang are with the College of Computer Science and Technology, Nanjing University of Aeronautics and Astronautics, China (e-mails: jinbo1608@nuaa.edu.cn; yangzhang@nuaa.edu.cn).

    J. Kang is with the School of Automation, Guangdong University of Technology, China (e-mail: kavinkang@gdut.edu.cn).

    D. Niyato is with the College of Computing and Data Science, Nanyang Technological University, Singapore (e-mail: dniyato@ntu.edu.sg).

    S. Mao is with the Department of Electrical and Computer Engineering, Auburn University, Auburn, AL, USA (e-mail: smao@ieee.org). 

\textit{Corresponding author: Yang Zhang.}
	}
}

\maketitle
	
\begin{abstract}
Industrial Cyber-Physical Systems (ICPSs) are an integral component of modern manufacturing and industries. By digitizing data throughout product life cycles, Digital Twins (DTs) in ICPSs enable a shift from current industrial infrastructures to intelligent and adaptive infrastructures. Thanks to data process capability, Generative Artificial Intelligence (GenAI) can drive the construction and update of DTs to improve predictive accuracy and prepare for diverse smart manufacturing. However, mechanisms that leverage Industrial Internet of Things (IIoT) devices to share sensing data for DT construction are susceptible to adverse selection problems. In this paper, we first develop a GenAI-driven DT architecture in ICPSs. To address the adverse selection problem caused by information asymmetry, we propose a contract theory model and develop a sustainable diffusion-based soft actor-critic algorithm to identify the optimal feasible contract. Specifically, we leverage dynamic structured pruning techniques to reduce parameter numbers of actor networks, allowing sustainability and efficient implementation of the proposed algorithm. Numerical results demonstrate the effectiveness of the proposed scheme and the algorithm, enabling efficient DT construction and updates to monitor and manage ICPSs.
\end{abstract}



\definecolor{limegreen}{rgb}{0.2, 0.8, 0.2}
\definecolor{forestgreen}{rgb}{0.13, 0.55, 0.13}
\definecolor{greenhtml}{rgb}{0.0, 0.5, 0.0}

\section{Introduction}\label{Intro}
With the advancement of industrial technologies, such as the Industrial Internet of Things (IIoT) and information communication technology, the convergence of physical and cyber spaces gives rise to a new paradigm called Industrial Cyber-Physical Systems (ICPSs) \cite{zhang2022advancements}. ICPSs are intricate and intelligent systems that integrate physical and computational components seamlessly, enabling real-time data exchange and decision-making\cite{10255287}. Predictive maintenance plays an increasingly pivotal role in the sustainability of ICPSs. The emergence of Artificial Intelligence (AI)-based approaches for predictive maintenance and monitoring has become prominent, such as transfer learning\cite{ZHANGtransfer} and Federated Learning (FL)\cite{ZHANGfederated}. To design integrated control and monitoring systems in ICPSs, Digital Twins (DTs) have received significant attention from academia and industry\cite{zhang2022advancements}. DTs refer to virtual replicas that cover the life cycle of physical entities. In ICPSs, through simulating the behavior and performance of industrial infrastructures based on real-time sensing data, DTs can offer predictive capabilities and provide insights to eliminate physical mistakes and attacks\cite{zhang2022advancements}, thus optimizing manufacturing processes and improving the performance of ICPSs. Nonetheless, the establishment and implementation of DTs in ICPSs depend on high-fidelity modeling and strong data interaction, which may meet data issues such as scarcity, bias, and noise.

Generative AI (GenAI) is a branch of AI technology that identifies the structures and patterns from existing data to generate various and original content\cite{wen2024generativeIoT}. The popular class of GenAI applications has emerged from foundation models, such as GPT-3 and Stable Diffusion, which are trained on vast quantities of data by leveraging different learning approaches. For example, ChatGPT is trained on a large corpus of text from diverse sources. This process enables it to acquire knowledge of linguistic patterns and structures, thus automatically generating valuable content based on the prompts provided by users. Relying on its incredible data processing and generation capabilities, GenAI technology has great potential to revolutionize various domains. For example, GenAI can drive the progression of modern IoT and enable more adaptive and intelligent IoT applications\cite{wen2024generativeIoT}, such as virtual assistants and smart surveillance.

GenAI also provides a novel tool for DT innovation \cite{chen2024generative, huang2024digital}. With the help of GenAI technology, the construction, maintenance, and optimization of DTs can be facilitated. In addition, researchers used GenAI technology to enhance DT emulation, feature abstraction, and decision-making modules\cite{huang2024digital}, driving innovation across diverse applications, especially ICPSs. For instance, in smart manufacturing, GenAI is capable of aiding the creation of adaptive DTs specifically designed for the manufacturing environment, which enhances efficiency in both production scheduling and real-time control. Although GenAI holds the potential to support emerging applications of DTs in ICPSs, there are challenges ahead:
\begin{itemize}
    \item \textbf{Challenge I:} Through extensive literature review, there is no research conducted on the utilization of GenAI to drive DTs within ICPSs. Hence, it is imperative to actively explore GenAI-driven DT architectures in ICPSs, thus improving system performance and sustainability.
    \item \textbf{Challenge II:} GenAI models can synthesize supplementary data based on real-world sensing data for DT construction. However, due to adverse selection problems caused by information asymmetry, IIoT devices may not be willing to share high-quality sensing data for DT construction without reliable incentives\cite{lotfi2023semantic}.
    \item \textbf{Challenge III:} It is a complex and high-dimensional problem to incentivize IIoT devices for DT construction\cite{lotfi2023semantic, zhang2022advancements}. Generative Diffusion Models (GDMs) as an advanced GenAI model show superior performance in solving high-dimensional decision-making problems\cite{du2024diffusion}, however, often at the cost of substantial computational overhead during training\cite{fang2024structural}. Therefore, it is necessary to develop sustainable GDM-based algorithms to solve high-dimensional optimization problems in ICPSs.
\end{itemize}

Some studies have been conducted to design incentive mechanisms for DT construction\cite{lotfi2023semantic,jiang2021cooperative}. However, there are no studies directly addressing the incentive mechanism design for DT construction in ICPSs. Moreover, GDMs for solving high-dimensional optimization problems still face substantial computational overhead because of the need for iterative denoising in generating new samples, and no work develops sustainable GDMs for optimal incentive design to mitigate the environmental impact of model training. To address Challenge I, we first design a GenAI-driven DT architecture in ICPSs. To address Challenge II, we propose a contract theory model to motivate IIoT devices to share sensing data for DT construction, where contract theory as an economic tool has been widely used to address incentive design problems with information asymmetry\cite{kang2023blockchain}. To address Challenge III, we develop a sustainable diffusion model-based algorithm to find the optimal feasible contract while mitigating the environmental impact for ICPSs. The contributions of this paper are summarized as follows:
\begin{itemize}
    \item We propose a GenAI-driven DT architecture in ICPSs. In particular, we systematically study how GenAI drives the DT construction pipeline in ICPSs, including the collection of real-time physical data, communications among DTs and between DTs and physical counterparts, DT modeling and maintenance, and DT decision-making, while also presenting the practical applications of these four parts (For Challenge I).
    \item To effectively alleviate the adverse selection problem caused by information asymmetry, we propose a contract theory model to motivate IIoT devices to participate in sensing data sharing for DT construction. Each IIoT device selects a suitable contract item that matches its type to obtain the highest utility, and the optimal feasible contract is achieved by maximizing the expected utility of the DT server (For Challenge II).
    \item To achieve sustainability for DT construction within ICPSs, we develop a sustainable diffusion-based Soft Actor-Critic (SAC) algorithm to generate the optimal contract under information asymmetry, where we apply dynamic structured pruning techniques to GDM-based networks for policy exploration in a more computationally efficient and scalable manner. \textit{To the best of our knowledge, this is the first work that leverages sustainable diffusion models for incentive design} (For Challenge III).
\end{itemize}

The rest of the paper is organized as follows. Section \ref{Related} reviews the related work. In Section \ref{architecture}, we introduce the GenAI-driven DT architecture in ICPSs. In Section \ref{contract}, we propose the contract model to incentivize IIoT devices to share sensing data for DT construction. In Section \ref{diffusion}, we develop the sustainable diffusion-based SAC algorithm to sustainably generate the optimal contract. In Section \ref{Results}, we present numerical results to demonstrate the effectiveness and sustainability of the proposed scheme and the algorithm. Section \ref{Conclusion} concludes the paper.

\section{Related Work}\label{Related}
In this section, we discuss several related works, including GenAI-driven DTs, incentive mechanism design for DT construction, and structured pruning techniques.

\subsection{Generative AI-driven Digital Twins}
GenAI possesses formidable capabilities to drive DTs from various domains\cite{bordukova2024generative, chen2024generative, huang2024digital, tao2024wireless}. In \cite{chen2024generative}, the authors explored the potential of GenAI-driven human DTs, including GenAI-enabled data acquisition, communication, data management, digital modeling, and data analytics. The authors in \cite{huang2024digital} proposed a GenAI-driven DT network architecture to realize intelligent external and internal closed-loop network management, where GenAI models can drive DT status emulation, feature abstraction, and network decision-making. In \cite{bordukova2024generative}, the authors introduced how GenAI facilitates DT modeling and provided existing implementations of GenAI for DTs in drug discovery and clinical trials. In \cite{tao2024wireless}, the authors explored the applications of GenAI models such as transformers and GDMs to empower DTs from several perspectives, including physical-digital modeling, synchronization, and slicing capability. While the integration of GenAI and DTs is capable of revolutionizing various sectors, there is no work systematically studying how GenAI can drive DTs in ICPSs.

\subsection{Incentive Mechanism Design for Digital Twin Construction}
Since DT construction requires high-quality data and intensive resources, incentive mechanisms play a crucial role in DT construction by promoting data sharing and resource collaboration. A few works have been conducted to design incentive mechanisms for DT construction \cite{lotfi2023semantic,jiang2021cooperative}. In \cite{lotfi2023semantic}, the authors proposed an iterative contract design to motivate IoT devices to share data for DT construction and used a multi-agent reinforcement learning algorithm to solve the formulated contract problem. In \cite{jiang2021cooperative}, the authors proposed a DT edge network framework for flexible and secure DT construction. To efficiently construct DTs, the authors also proposed an iterative double auction-based joint cooperative FL and local model update verification scheme. Incentive mechanism design for ensuring high-quality DT construction is critical, and this topic is still worth investigating, especially in ICPSs. However, none of the existing studies have used GenAI models to discover the optimal incentive design for DT construction.


\begin{figure*}[t]
\centerline{\includegraphics[width=0.97\textwidth]{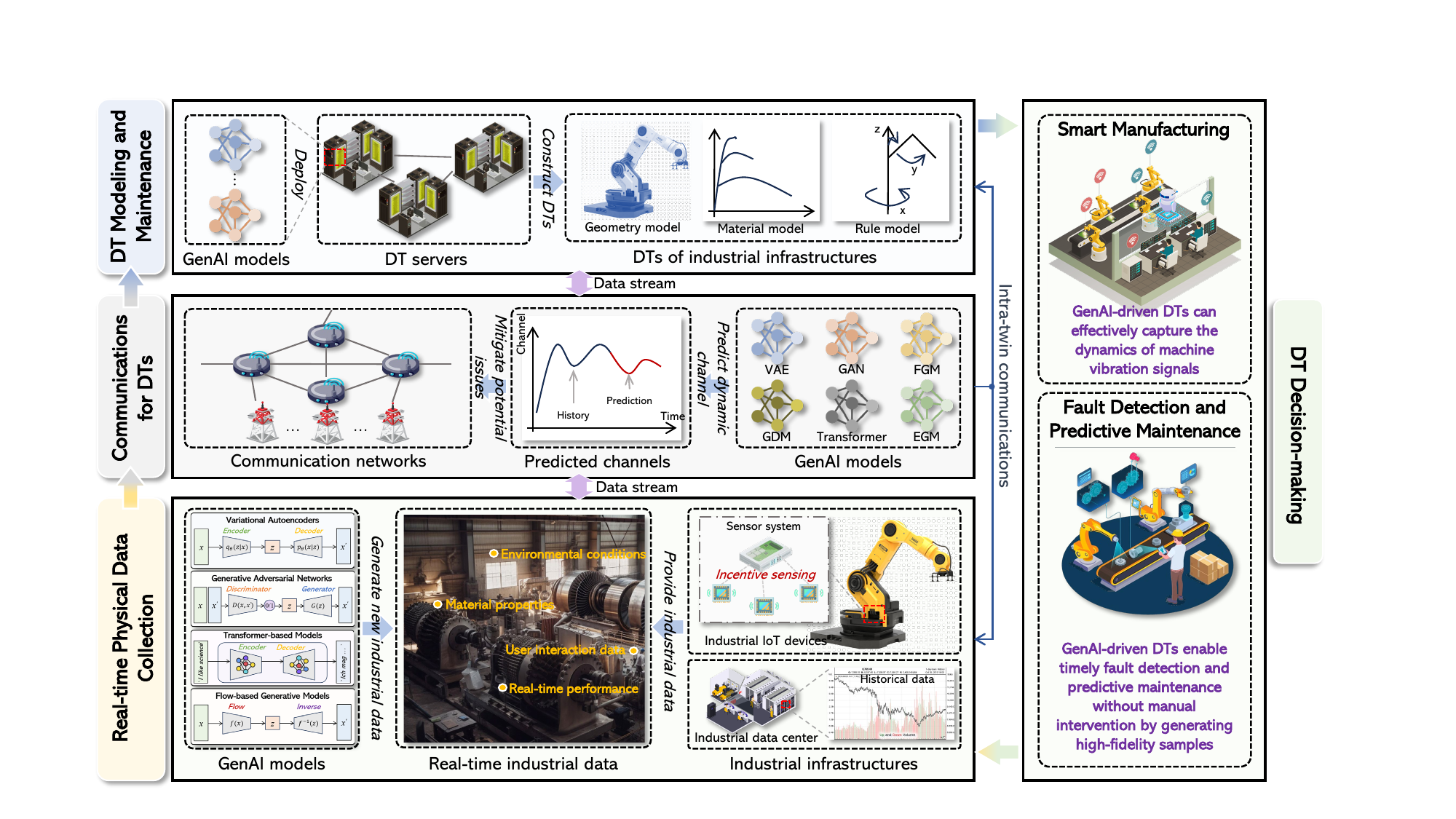}}
\caption{The architecture of GenAI-driven DTs in ICPSs. We study how GenAI drives the DT construction pipeline, i.e., real-time physical data collection, communications for DTs, DT modeling and maintenance, and DT decision-making.}
\label{Digital}
\end{figure*}

\subsection{Structured Pruning Techniques}
Structured pruning techniques focus on eliminating parameters and substructures from networks\cite{fang2024structural}, which has been used to effectively optimize Deep Reinforcement Learning (DRL) algorithms. For example, the authors in \cite{Kangtiny} proposed a tiny multi-agent DRL algorithm to obtain the optimal game solution, which leverages the structured pruning technique to reduce the parameter number of the actor-critic network. The authors in \cite{su2024compressing} proposed a dynamic structured pruning approach to remove the unimportant neurons of DRL models during the training stage. In \cite{fang2024structural}, the authors proposed a dedicated method for compressing diffusion models by utilizing Taylor expansion. Diffusion model-based DRL algorithms have been proven to be an effective method to solve network optimization problems\cite{10638123, du2024diffusion}. This algorithm uses multi-layer perceptions for iterative denoising from the Gaussian noise\cite{du2024diffusion}, which consumes a lot of energy. However, there is no research on developing efficient diffusion model-based DRL algorithms for network optimization. Therefore, we, for the first time, apply structured pruning techniques to diffusion model-based DRL algorithms for optimal incentive design, making them more efficient and sustainable.

\section{Generative AI-driven Digital Twin Architecture}\label{architecture}
In this section, we introduce the proposed GenAI-driven DT architecture in ICPSs, which systematically studies the role of GenAI on the DT construction pipeline, as shown in Fig. \ref{Digital}.

\subsection{Real-time Physical Data Collection}
Real-time data collection from the physical space is the first step in constructing DTs in ICPSs\cite{yu2024bi}. The required data, such as historical equipment parameters and real-time operation, are collected by industrial infrastructures such as IIoT devices and the industrial data center\cite{zhang2022advancements}, where real-time sensing data allows DTs to continuously mirror the current state of the physical system, and historical data can be used for training predictive models within DTs. However, due to the heterogeneity of diverse sensing sources, some real-world data may be insufficient and confidential\cite{zhang2022advancements}, affecting the rendering quality of DTs. Fortunately, GenAI supports data generation by capturing data distributions from the actual physical system\cite{wen2024generativeIoT}, which can provide ultra-realistic industrial data based on the collected data to improve DT training. For instance, the authors in \cite{tao2024wireless} utilized a Generative Adversarial Network (GAN)-based model to generate a sufficient dataset, which is used to train DT models.

\subsection{Communications for Digital Twins}
The key to DT construction is to send the collected data in real time to the server where DTs are deployed \cite{yu2024bi}. Communications among DTs and between DTs and physical counterparts act as a bridge for data transmission between physical and virtual spaces\cite{chen2024generative}, such as collected data transmission and DT feedback. However, due to the potentially harsh industrial environments, the communication channel conditions in ICPSs may dynamically change\cite{zhang2022advancements}, causing reliability issues in DT communications. Predictive DT channels provide an effective solution that enables DTs to maintain high fidelity in real-time communications\cite{10417075}. Fortunately, GenAI can generate synthetic data representing various channel conditions, and the generated data and raw data can be used together to train deep learning models for predicting the DT communication channel dynamics, thus mitigating potential issues such as communication disruptions and ensuring efficient and reliable DT communications. For example, a recent study in \cite{10417075} proposed a convolutional time-series GAN to generate synthetic data based on original channel data, which can enhance the accuracy of ConvLSTM-based predictive DT channels in dynamic channel conditions.

\subsection{Digital Twin Modeling and Maintenance}
Based on collected data, high-fidelity DTs can be constructed by reproducing the geometry and physical properties of industrial infrastructures\cite{yu2024bi}. These virtual models are maintained on edge servers to ensure real-time updates\cite{yu2024bi}. However, due to the complicated environments of ICPSs\cite{zhang2022advancements}, traditional DT modeling methods, such as structural modeling or behavioral modeling\cite{yu2024bi}, rely on accurate simulation parameters\cite{chen2024generative}, making it challenging to continuously guarantee the quality of DTs. Thanks to the robust data generation capability, GenAI can provide supplementary data for DT modeling in ICPSs, including environmental conditions and operational parameters, and even allow DTs to update and refine their knowledge base over time. For instance, the authors in \cite{HU2023103807} developed a Wasserstein GAN-based DT in industrial asset health monitoring. Besides, the augmented data can preserve almost all patterns from the original data\cite{chen2024generative}, thus protecting private data in ICPSs.

\subsection{Digital Twin Decision-Making}
After achieving the virtual-real mapping, DTs can timely analyze real-world data and provide decision-making results to ICPSs \cite{yu2024bi}. GenAI-driven DTs enable intelligent and reliable applications, such as smart manufacturing\cite{wen2024generativeIoT}, intelligent assessment\cite{FENG2023109896}, and fault detection and predictive maintenance\cite{10195218, FENG2022108319}. For example, in smart manufacturing\cite{wen2024generativeIoT}, GenAI-driven DTs can forecast product life in a timely manner and evaluate manufacturing plans based on generated fault diagnosis. Then, manufacturers can react according to the feedback conveyed by DTs. In the intelligent assessment of gear surface degradation\cite{FENG2023109896}, by generating synthetic data that represents surface defects, GenAI-driven DTs can recognize different stages and types of gear wear with higher precision. Besides, in fault detection and predictive maintenance\cite{10195218}, by combining GenAI and domain adversarial graph networks, DTs can better diagnose bearing faults by learning domain-invariant features.

\section{Contract Modeling}\label{contract}
In ICPSs, IIoT devices equipped with a set of sensors can collect geospatial data from the surrounding environment and send the collected data to DT servers for DT construction\cite{lotfi2023semantic}. If the data quality is low, the created DTs may not accurately reflect the real-time dynamics of industrial infrastructures.
Thus, high-quality sensing data is critical for DT construction. However, due to the high cost of collecting sensing data in ICPSs\cite{zhang2022advancements}, IIoT devices may not actively provide high-quality sensing data to DT servers, or even maliciously provide harmful data\cite{kang2023blockchain}, leading to adverse selection problems caused by information asymmetry\cite{lotfi2023semantic}. To this end, we use contract theory to motivate IIoT devices to provide high-quality sensing data for DT construction, where contract theory as a powerful economy tool aims at addressing information asymmetry\cite{kang2023blockchain, lotfi2023semantic}. In this paper, we consider a DT server and $M$ IIoT devices for DT construction. 

\subsection{Utility of IIoT Devices}
We denote $\Psi = \{\psi_k|\psi_k\in \mathbb{N}_+, \:k \in \small\{1,\ldots, K,\: K \leq M\small\}\}$ as the set of different sensing and communication levels of IIoT devices\cite{lotfi2023semantic}, where the level is mainly impacted by equipment and algorithms used by IIoT devices for data sensing and transmission\cite{wen2024surveyintegratedsensingcommunication}. While the sensing and communication levels of IIoT devices are not fully disclosed due to information asymmetry\cite{lotfi2023semantic}, the DT server can leverage statistical distributions of IIoT devices from historical data to group IIoT devices into discrete types\cite{kang2023blockchain}. Thus, we consider sensing and communication levels as types and sort the types in ascending order, i.e., $0 < \psi_1 \leq \cdots \leq \psi_K$, where $K$ is the number of types of IIoT devices. Especially, IIoT devices with higher sensing and communication levels are characterized as higher types, which can provide better sensing data. Based on \cite{lotfi2023semantic}, the utility of an IIoT device with type $\psi_k$ to send sensing data with size $\hat{s}_k$ to the DT server is defined as
\begin{equation}
    \Tilde{u}_k(\hat{s}_k, r_k) = \rho \psi_k r_k - c\hat{s}_k - c_0,
\end{equation}
where $\rho$ is a pre-defined weight parameter about the incentive, $r_k$ is the received reward associated with the provided data $\hat{s}_k$, $c$ is the unit cost related to the collection, computation, and transmission of sensing data $\hat{s}_k$\cite{kang2023blockchain}, and $c_0$ is an additional cost involving energy consumption\cite{lotfi2023semantic}.

\subsection{Utility of Digital Twin Server}
Considering the more high-quality data provided by IIoT devices, the higher satisfaction of the DT server, we adopt the $\beta$-fairness function $g(\hat{s}_k)$ to quantify the satisfaction of the DT server\cite{lotfi2023semantic}, given by
\begin{equation}
    g(\hat{s}_k) = \frac{1}{1-\beta}\hat{s}_k^{1-\beta},
\end{equation}
where $0 \leq \beta < 1$ is a pre-defined constant. Since the DT server just knows the number of IIoT devices and the distribution of each type due to information asymmetry, the overall utility of the DT server is given by
\begin{equation}
    U(\bm{\hat{s}}, \bm{r}) = M\sum_{k=1}^Kq_k(\vartheta g(\hat{s}_k) - r_k),
\end{equation}
where $\vartheta > 0$ is the unit revenue for the satisfaction of the DT server\cite{kang2023blockchain}, $q_k$ represents the probability that IIoT devices belong to type $\psi_k$, and $\bm{\hat{s}} = [\hat{s}_k]_{1\times K}$ and $\bm{r} = [r_k]_{1\times K}$ represent the vectors of high-quality data volume sizes and rewards, respectively.

\subsection{Contract Formulation}
To maximize the overall utility, the DT server designs a contract consisting of a group of contract items. The contract is denoted as $\varOmega = \{(\hat{s}_k, r_k),k \in \{1, \ldots, K \}\}$, where the more $\hat{s}_k$, the higher $r_k$. To ensure that each IIoT device selects the most suitable contract item to maximize its utility, the feasible contract should satisfy the Individual Rationality (IR) and Incentive Compatibility (IC) constraints\cite{kang2023blockchain, lotfi2023semantic}:
\begin{definition}
\textbf{Individual Rationality:} Type-$k$ IIoT devices gain non-negative utilities by selecting the contract item $(\hat{s}_k, r_k)$ that matches their types, i.e.,
\begin{equation}\label{IR}
    \Tilde{u}_k(\hat{s}_k, r_k) = \rho\psi_k r_k - c\hat{s}_k - c_0 \geq 0,\: \forall k \in \small\{1,\ldots,K\small\},
\end{equation}
\end{definition}
\begin{definition}
\textbf{Incentive Compatibility:} Any type of IIoT device prefers to choose the contract item $(\hat{s}_k, r_k)$ designed for it rather than any other contract item $(\hat{s}_n, r_n),\: n\in \small\{1,\ldots,K\small\}$, and $n\neq k$, i.e.,
\begin{equation}\label{IC}
        \rho\psi_k r_k - c\hat{s}_k \geq \rho\psi_k r_n - c\hat{s}_n,\:\forall k,n\in\small\{1,\ldots,K\small\},\: k \neq n.
\end{equation}
\end{definition}
The IR constraints (\ref{IR}) guarantee that the utility of IIoT devices is non-negative, and the IC constraints (\ref{IC}) guarantee that each IIoT device can achieve the highest utility by selecting the optimal contract item designed for its type. The DT server aims to maximize its overall utility. Based on the above IC and IR constraints, the optimization problem is expressed as
\begin{equation}\label{problem1}
    \begin{split}
        &\max\limits_{\bm{\hat{s}},\bm{r}}\:U (\bm{\hat{s}},\bm{r}) \\
        &\:\:\mathrm{s.t.}\:\:\Tilde{u}_k(\hat{s}_k, r_k) \geq 0,\: \forall k \in \small\{1,\ldots,K\small\},\\
        &\qquad\: \Tilde{u}_k(\hat{s}_k, r_k) \geq \Tilde{u}_k(\hat{s}_n, r_n),\: \forall n, k \in \small\{1,\ldots,K\small\},\: n\neq k,\\
        &\qquad\:  \hat{s}_k \geq 0, r_k \geq 0, \psi_k > 0,\: \forall k \in \small\{1,\ldots,K\small\}.
    \end{split}
\end{equation}

\begin{remark}
In the optimization problem (\ref{problem1}), there are $K$ IR constraints and $K(K-1)$ IC constraints, which are all non-convex. Thus, it is difficult to directly solve the problem (\ref{problem1}). The standard approach is to reduce IR and IC constraints by defining some lemmas and use heuristic and traditional optimization algorithms to solve the derived problem\cite{10271832}. However, these approaches add difficulty to the problem formulation, which is time-consuming to adjust and prove the corresponding lemmas\cite{lotfi2023semantic}. Besides, states of IIoT devices and the DT server change dynamically due to the heterogeneity and dynamics of ICPSs\cite{zhang2022advancements}, and traditional approaches need to be re-designed in practice. Furthermore, these approaches may not be applicable in practical scenarios when the environment lacks complete and accurate information. Thus, traditional mathematical techniques may not effectively solve the problem (\ref{problem1}) in practice. DRL algorithms have been widely used to solve optimization problems because they can learn complex policies in high-dimensional spaces, which can find near-optimal solutions to challenging problems where traditional methods might struggle. In particular, GDMs possess a robust generative capability and have been used to enhance DRL algorithms, thus being able to find an optimal policy for dynamic and high-dimensional optimization problems\cite{du2024diffusion}. For efficient implementation in ICPSs, we propose a sustainable GDM-based DRL algorithm to solve (\ref{problem1}).
\end{remark}

\section{Sustainable Diffusion-based Soft Actor-Critic Algorithms for Optimal Contract Design}\label{diffusion}
\begin{figure}[t]
    \centering  
    \includegraphics[width=0.47\textwidth]{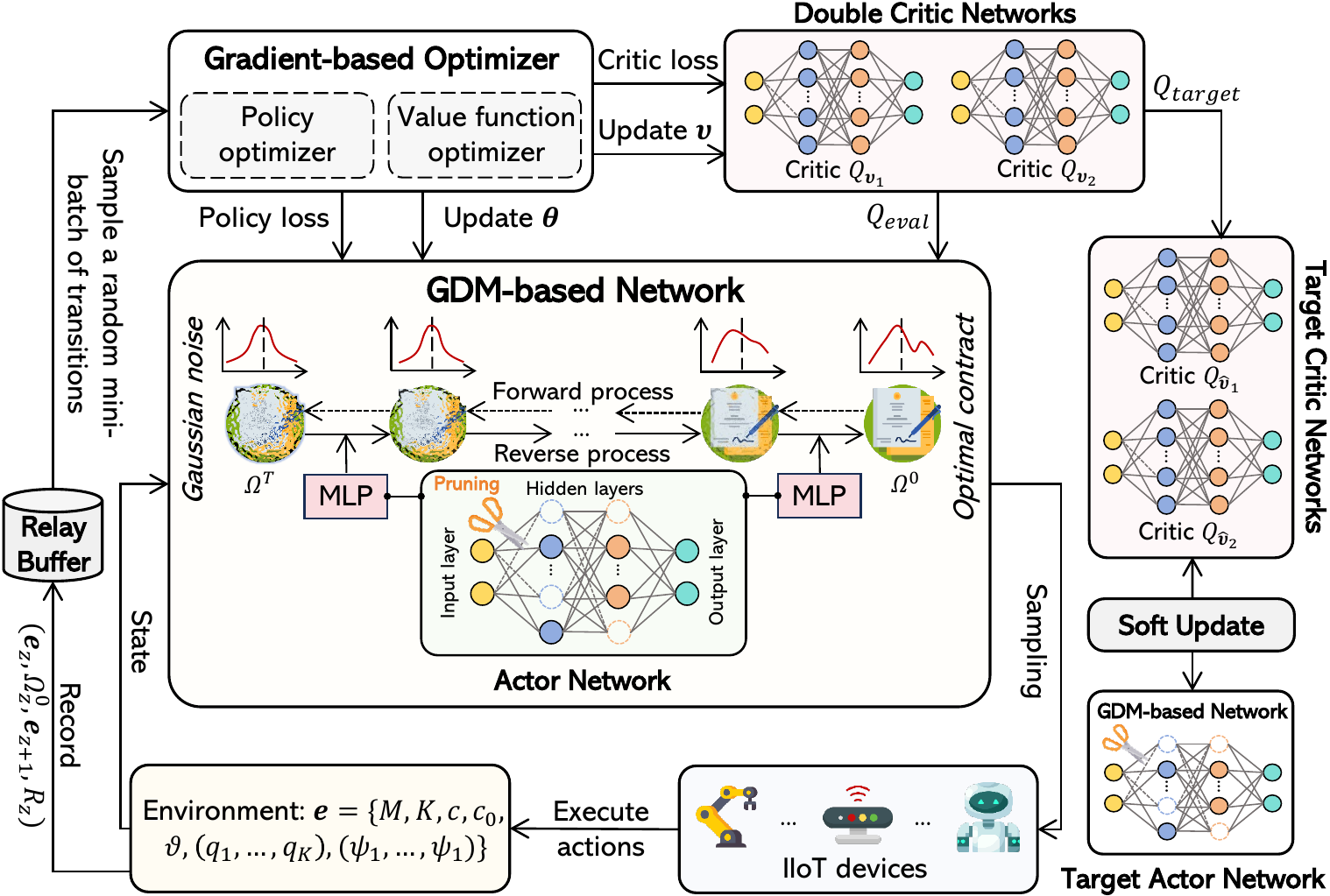}   
    \caption{The proposed algorithm architecture, where we utilize dynamic structured pruning techniques to sparsify the actor networks of the diffusion model. The goal of the algorithm is to design optimal contracts that motivate IIoT devices to provide sensing industrial data, which is one of the data sources for DT construction, as shown in Fig. \ref{Digital}.}
    \label{Diff_aglorithm}
\end{figure}

In this section, we propose the sustainable diffusion-based SAC algorithm to efficiently generate the optimal feasible contract under information asymmetry, where we apply dynamic structured pruning techniques to GDM-based networks to reduce the training cost and carbon emissions. The architecture of the proposed algorithm is shown in Fig. \ref{Diff_aglorithm}.

\subsection{Generative Diffusion Models for Optimal Contract Design}
Diffusion models work by corrupting the training data by continuously adding Gaussian noise and then learning to recover the data by reversing this noise process\cite{fang2024structural}. Through an iterative process of denoising the initial distribution, GDMs can generate the optimal contract $\varOmega^0 = {(\hat{s}_k^*, r_k^*), k \in \{1, \ldots, K \}}$. During iterations of $T$, Gaussian noise is gradually added to the initial contract $\varOmega$, emerging a series of contract samples $(\varOmega^1, \varOmega^2,\ldots,\varOmega^T)$. We define the environment during the optimal contract design as
\begin{equation}
    \bm{e} \triangleq \small\{M,K,c,c_0,\vartheta,(q_1,\ldots,q_K),(\psi_1, \ldots, \psi_K)\small\}.
\end{equation}

We denote the contract design policy as $\pi_{\bm{\theta}}(\varOmega|\bm{e})$ with parameters $\bm{\theta}$, which is constituted by the GDM-based network, as shown in Fig. \ref{Diff_aglorithm}. Through the reverse process of a conditional diffusion model, the policy $\pi_{\bm{\theta}}(\varOmega|\bm{e})$ can be expressed as\cite{wang2022diffusion}
\begin{equation}
    \pi_{\bm{\theta}}(\varOmega|\bm{e})=p_{\bm{\theta}}(\varOmega^{0:T}|\bm{e})=\mathcal{N}(\varOmega^T;\textbf{0},\textbf{I})\prod_{t=1}^Tp_{\bm{\theta}}(\varOmega^{t-1}|\varOmega^t,\bm{e}),
\end{equation}
where the end sample of the reverse chain is the selected contract $\varOmega^0$. $p_{\bm{\theta}}(\varOmega^{t-1}|\varOmega^t,\bm{e})$ as a noise prediction model is modeled as a Gaussian distribution $\mathcal{N}(\varOmega^{t-1};\boldsymbol{\mu}_{\bm{\theta}}(\varOmega^t,\bm{e},t),\boldsymbol{\Sigma}_{\bm{\theta}}(\varOmega^t,\bm{e},t))$ with the covariance matrix expressed as $\boldsymbol{\Sigma}_{\bm{\theta}}(\varOmega^t,\bm{e},t)=\delta_t\textbf{I}$\cite{wang2022diffusion}, and the mean $\boldsymbol{\mu}_{\bm{\theta}}(\varOmega^t,\bm{e},t)$ is given by
\begin{equation}\label{mean}
    \boldsymbol{\mu}_{\bm{\theta}}(\varOmega^t,\bm{e},t)=\frac{1}{\sqrt{\alpha_t}}\Big(\varOmega^t-\frac{\delta_t}{\sqrt{1-\bar{\alpha_t}}}\boldsymbol{\epsilon}_{\bm{\theta}}(\varOmega^t,\bm{e},t)\Big),
\end{equation}
where $\delta_t \in (0,1)$ is a hyperparameter for model training, $\alpha_t = 1 - \delta_t$, and $\Bar{\alpha_t} = \prod_{j=0}^t \delta_j$\cite{wang2022diffusion,du2024diffusion}. $\boldsymbol{\epsilon}_{\bm{\theta}}$ is a deep model that generates contracts conditioned on the environment $\bm{e}$, as determined by the policy. We first sample $\varOmega^T \sim \mathcal N (\textbf{0}, \textbf{I})$ and then the selected contract can be sampled via the reverse diffusion chain parameterized by $\bm{\theta}$, given by
\begin{equation}\label{denoise}
    \varOmega^{t-1}|\varOmega^t=\frac{\varOmega^t}{\sqrt{\alpha_t}}-\frac{\delta_t}{\sqrt{\alpha_t(1-\bar{\alpha_t})}}\boldsymbol{\epsilon}_{\bm{\theta}}(\varOmega^t,\bm{e},t)+\sqrt{\delta_t}\boldsymbol{\epsilon}.
\end{equation}

To improve the policy $\pi_{\bm{\theta}}(\varOmega|\bm{e})$, we introduce the Q-function\cite{wang2022diffusion}, which can guide the reverse diffusion chain to preferentially sample contracts with high values. We build two critic networks $Q_{\bm{\upsilon}_1}, Q_{\bm{\upsilon}_2}$, target critic networks $Q_{\hat{\bm{\upsilon}}_1}, Q_{\hat{\bm{\upsilon}}_2}$, and target policy $\hat{\pi}_{\hat{\bm{\theta}}}$. Then, we define the contract quality network as $Q_{\bm{\upsilon}}(\bm{e}, \varOmega) = \min\small\{Q_{\bm{\upsilon}_1}(\bm{e}, \varOmega), Q_{\bm{\upsilon}_2}(\bm{e}, \varOmega)\small\}$. Thus, the optimal contract design policy aims at maximizing the expected cumulative reward, expressed as\cite{du2024diffusion}
\begin{equation}
    \pi = \arg\max_{\pi_{\bm{\theta}}} \Bbb E \Bigg[\sum_{z = 0} ^ Z \gamma^z (R(\bm{e}_z,\varOmega_z)+\varsigma H(\pi_{\bm{\theta}}(\bm{e}_z)))\Bigg],
\end{equation}
where $Z$ denotes the maximum training step, $\gamma$ denotes the discount factor for future rewards, $R(\bm{e}_z,\varOmega_z)$ represents the immediate reward upon executing action $\varOmega_z$ in state $\bm{e}_z$, $\varsigma$ is the temperature coefficient controlling the strength of the entropy, and $H(\pi_{\bm{\theta}}(\bm{e}_z))=-\pi_{\bm{\theta}}(\bm{e}_z)\log\pi_{\bm{\theta}}(\bm{e}_z)$ is the entropy of the policy $\pi_{\bm{\theta}}(\bm{e}_z)$\cite{haarnoja2018soft}. To update the Q-function, the optimization of $\bm{\upsilon}_m$ for $m=\small\{1,2\small\}$ aims at minimizing the objective function as \cite{wang2022diffusion}
\begin{equation}\label{Q_update}
\begin{split}
    &\Bbb E_{(\bm{e}_z, \varOmega_z, \bm{e}_{z+1}, R_z)\sim \mathcal{B}_z, \Phi^0_{z+1}\sim\hat{\pi}_{\hat{\bm{\theta}}}}\Big[\sum_{m=1,2}(R(\bm{e}_z,\varOmega_z)\\
    &\:+\gamma^z(1-d_{z+1})\hat{\pi}_{\hat{\bm{\theta}}}(\bm{e}_{z+1})Q_{\hat{\bm{\upsilon}}}(\bm{e}_{z+1})-Q_{\bm{\upsilon}_m}(\bm{e}_z,\varOmega_z))^2\Big],
\end{split}
\end{equation}
where $\mathcal{B}_z$ is a mini-batch of transitions with a size $b$ sampled from the experience replay memory $\mathcal{D}$ and $d_{z+1}$ is a $0$-$1$ variable that represents the terminated flag\cite{du2024diffusion}.

\subsection{Dynamic Structured Pruning}
In terms of the network structure, both the policy and critic networks are multilayer perception networks, which consist of an input layer, multiple hidden layers, and an output layer. Considering a $L$-layer policy network, where parameters in the $l$-th fully-connected layer are denoted by $\bm{\theta}^{(l)},\: l \in \{1, \ldots, L \}$, the output of the $l$-th layer is expressed as 
\begin{equation}
    \bm{h}^{(l)} = f^{(l)}\big(\bm{\theta}^{(l)}\bm{h}^{(l-1)} + b^{(l)}\big),
\end{equation}
where $f^{(l)}$ is the activation function of the $l$-th layer and $b^{(l)}$ is the deviation at the $l$-th layer. We focus on pruning the redundant neurons and connected weights of policy networks without affecting the performance, which enhances the training efficiency of the policy network. To this end, we introduce a binary mask $m^{(l)}_i$ to represent the pruning state of each neuron $o_i^{(l)}$ at the $l$-th layer\cite{su2024compressing}, where $m^{(l)}_i = 0$ indicates that the neuron $o_i^{(l)}$ should be pruned, and $m^{(l)}_i = 1$ indicates that the neuron $o_i^{(l)}$ should be reserved. Based on the binarized mask $\bm{m}^{(l)}$, the output of the $l$-th layer is rewritten as
\begin{equation}
    \bm{h}^{(l)} = f^{(l)}\big(\bm{\theta}^{(l)}\bm{h}^{(l-1)}\odot \bm{m}^{(l)} + b^{(l)}\big),
\end{equation}
where $\odot$ represents the Hadamard product. We use the policy gradient algorithm to update the contract design policy\cite{du2024diffusion}. Specifically, the policy gradient with respect to the policy parameters $\bm{\theta}$ can be computed as the expectation over $\mathcal{B}_z$. Thus, the gradient is given by
\begin{equation}\label{gradient}
    \nabla_{\bm{\theta}_z}J(\bm{\theta}) = \Bbb E_{\bm{e}_z\sim \mathcal{B}_z}[-\nabla_{\bm{\theta}_z}\pi_{\bm{\theta}_z}(\bm{e}_z)Q_{\bm{\upsilon}_z}(\bm{e}_z)],
\end{equation}
where $\bm{\theta}_z$ and $\bm{\upsilon}_z$ are the policy and critic parameters at the $z$-th training step, respectively. Thus, the policy parameters $\bm{\theta}$ are updated by performing gradient descent based on (\ref{gradient}), which is expressed as\cite{du2024diffusion,su2024compressing}
\begin{equation}\label{policy_update}
    \bm{\theta}_{z+1}^{(l)} \gets \bm{\theta}_{z}^{(l)}-\eta \nabla_{\bm{h}_z^{(l)}\odot\bm{m}_z^{(l)}}J(\bm{\theta})\nabla_{\bm{\theta}_z^{(l)}}(\bm{h}_z^{(l)}\odot\bm{m}_z^{(l)}),
\end{equation}
where $\eta\in (0,1]$ is the learning rate of the actor network. For the target networks that have the same network structure as the online networks\cite{du2024diffusion}, the parameters of the target policy are also performed dynamic structured pruning, and their parameters are updated by using a soft update mechanism, given by\cite{wang2022diffusion} 
\begin{equation}\label{targets}
\begin{split}
    \hat{\bm{\theta}}_{z+1} &\gets \varepsilon \bm{\theta}_z + (1-\varepsilon)\hat{\bm{\theta}}_{z},\\
    \hat{\bm{\upsilon}}_{m, z+1} &\gets \varepsilon \bm{\upsilon}_{m, z} + (1-\varepsilon)\hat{\bm{\upsilon}}_{m, z},\: \text{for}\: m = \small\{1,2\small\},
\end{split}  
\end{equation}
where $\varepsilon\in(0,1]$ is the update rate of the target network.

The pruning threshold plays an important role in the pruning decisions of network parameters\cite{Kangtiny, su2024compressing}. We adopt a dynamic pruning threshold $\varUpsilon$, which is given by\cite{Kangtiny}
\begin{equation}\label{dynamic_th}
\begin{split}
    \varUpsilon_z &= \sum_{l = 1}^L\sum_{i=1}^I \phi_i^{(l)}\omega_z,\\
    \omega_z &= \hat{\omega} - \hat{\omega}\Big(1-\frac{z}{N\triangle}\Big)^3,
\end{split}
\end{equation}
where $\phi_i^{(l)}$ and $N$ represent the importance of the $i$-th neuron of the $l$-th layer and the total number of pruning steps, respectively. $\triangle$ denotes the pruning frequency, and $\omega_z$ and $\hat{\omega}$ represent the current sparsity at the $z$-th training step and the target sparsity, respectively.

After determining the pruning threshold, neurons ranked below the threshold will be pruned\cite{Kangtiny, su2024compressing}, and the binary masks used for pruning are updated, given by\cite{Kangtiny}
\begin{equation}\label{mask}
    m_i^{(l)} = 
    \begin{cases}
        1, &\text{if} \: |m_i^{(l)} \theta_i^{(l)}|\geq\varUpsilon , \\
        0, &\text{otherwise}.
    \end{cases}
\end{equation}

Finally, we build a compact policy network according to the redundancy of the sparse policy network. The procedure of the proposed sustainable diffusion model is shown in Algorithm \ref{diffusion_algorithm}. The computational complexity of the proposed algorithm consists of three parts. In the initialization part, the computational complexity is $\mathcal{O}(|\bm{\theta}|+|\bm{\upsilon}|)$. In the training part, the computational complexity is $\mathcal{O}(Z (T\sum_{l=1}^{L-1}|\bm{\theta}^{(l)}| + \sum_{l=1}^{L-1}|\bm{h}^{(l)}|))$\cite{du2024diffusion,Kangtiny}. In the inference part, the computational complexity is $\mathcal{O}(|\hat{\bm{\theta}}|)$. Thus, the computational complexity of the proposed algorithm is around $\mathcal{O}(|\bm{\theta}|+|\bm{\upsilon}|+|\hat{\bm{\theta}}|+Z (T\sum_{l=1}^{L-1}|\bm{\theta}^{(l)}| + \sum_{l=1}^{L-1}|\bm{h}^{(l)}|))$, where $T$ is the total number of denosing steps. In the following, we compare the proposed sustainable diffusion-based SAC algorithm with DRL algorithms that are commonly used for optimal contract design to evaluate its performance, i.e., Proximal Policy Optimization (PPO), SAC, diffusion-based SAC, and diffusion-based Deep Deterministic Policy Gradient (DDPG). We simply introduce these four algorithms and present their computational complexities for solving the optimization problem (\ref{problem1}). Specifically,
\begin{itemize}
    \item \textit{PPO}: Update the policy by using a clipped surrogate objective\cite{schulman2017proximalpolicyoptimizationalgorithms}. Its computational complexity is $\mathcal{O}\big(\sum_{j=1}^J o_{j-1}o_j\big)$\cite{10032267}, where $o_j$ represents the number of neural units at the $j$-th layer in the policy network.
    \item \textit{SAC}: Optimize the stochastic policy in an off-policy way, which maximizes a trade-off between expected return and entropy\cite{haarnoja2018soft}. Its computational complexity is $\mathcal{O}\left(\sum_{j=1}^{J} o_{j-1}^{Q} o_{j}^{Q} + \sum_{i=1}^{I} o_{i-1}^{\pi} o_{i}^{\pi}\right)$\cite{haarnoja2018soft}, where $o_{j}^{Q}$ and $o_{i}^{\pi}$ represent the number of neuron units at the $j$-th layer of $Q_{\bm{\upsilon}}(\bm{e}, \varOmega)$ and the ${i}$-th layer of $\pi_{\bm{\theta}}(\varOmega|\bm{e})$, respectively.
    \item \textit{Diffusion-based SAC}: Use diffusion models to enhance SAC algorithms\cite{du2024diffusion}. Its computational complexity is $\mathcal{O}\left(Z[CV+TC|\bm{\theta}|+(B+1)(|\bm{\theta}|+|\bm{\upsilon}|)]\right)$\cite{du2024diffusion}, where $C$ represents the maximum number of collected transitions at each training step, $V$ represents the complexity of interacting with the environment $\bm{e}$, and $B$ represents the batch size.
    \item \textit{Diffusion-based DDPG}: Use diffusion models to enhance DDPG algorithms that learn the policy through the Q-learning function\cite{10707303}. Its computational complexity is $\mathcal{O}\left(Z[T|\bm{\theta}|+|\bm{\upsilon}|]\right)$\cite{10707303}.
\end{itemize}

\begin{algorithm}[t]
\label{diffusion_algorithm}
\DontPrintSemicolon
\SetAlgoLined
\KwIn {Environment $\bm{e}$ and diffusion parameters.}
\KwOut {The optimal contract design $\varOmega^0$.}
\#\# \textbf{\textit{Initialize}}

Initialize relay buffer $\mathcal{D}$, policy network $\pi_{\bm{\theta}}$, critic networks $Q_{\bm{\upsilon}_1}, Q_{\bm{\upsilon}_2}$, target networks $\hat{\pi}_{\hat{\bm{\theta}}}, Q_{\hat{\bm{\upsilon}}_1}, Q_{\hat{\bm{\upsilon}}_2}$, and binary masks $\bm{m}$.

\#\# \textbf{\textit{Training}}

\For{ \rm{the training step} $z=1$ \rm{to} $Z$}
{
\#\#\# \textit{Generating contracts}

    Observe environment $\bm{e}_z$ and initialize a random process $\mathcal{N}$ for contract design exploration.

    Set $\varOmega_z^T$ as Gaussian noise and generate contract design $\varOmega_z^0$ by  (\ref{denoise}).

    Execute $\varOmega_z^0$, observe the next environment $\bm{e}_{z+1}$, and obtain reward $R_z$. 

    Store record $(\bm{e}_z,\varOmega_z^0,\bm{e}_{z+1}, R_z)$ into $\mathcal{D}$.

    Sample a random mini-batch of transitions $\mathcal{B}_z$ from replay buffer $\mathcal{D}$.

    Update $Q_{\bm{\upsilon}_1}, Q_{\bm{\upsilon}_2}$ using $\mathcal{B}_z$ by (\ref{Q_update}).

    \#\#\# \textit{Dynamic pruning and fine-tuning}
    
    Compute neuron importance $\bm{\phi}_z$ based on \cite{su2024compressing}.
    
    Update the policy $\pi_{\bm{\theta}}$ using $\mathcal{B}_z$ by (\ref{policy_update}).

    Update the target networks $\hat{\pi}_{\hat{\bm{\theta}}}, Q_{\hat{\bm{\upsilon}}_1}, Q_{\hat{\bm{\upsilon}}_2}$ by (\ref{targets}).

    Compute the dynamic pruning threshold $\varUpsilon$ for the policy network parameters $\bm{\theta}_z$ by (\ref{dynamic_th}).
    
    Update binary masks $\bm{m}_z$ for $\bm{\theta}_z$ by (\ref{mask}).
    
    \If{$\phi_i^{(l)} < \varUpsilon$} {
        Remove the $i$-th neuron at the $l$-th layer with the mask $m_i^{(l)}$ and associated parameters $\bm{\theta}_z^{(l)}$ from the policy network.\;
    }
}
Reconstruct the compact policy networks.

\#\# \textbf{\textit{Inference}}

    Input the environment vector $\bm{e}$.
    
    Generate the optimal contract design $\varOmega^0$ based on the target policy $\hat{\pi}_{\hat{\bm{\theta}}}$ by (\ref{denoise}).
    
    \textbf{return} $\varOmega^0=\small\{(\hat{s}_k^*, r_k^*), 1 \leq k \leq K \small\}$.
\caption{Sustainable Diffusion Models based on Dynamic Structured Pruning}\label{diffusion_algorithm}
\end{algorithm}

\begin{figure}[!t]
    \centering  
    \includegraphics[width=0.46\textwidth]{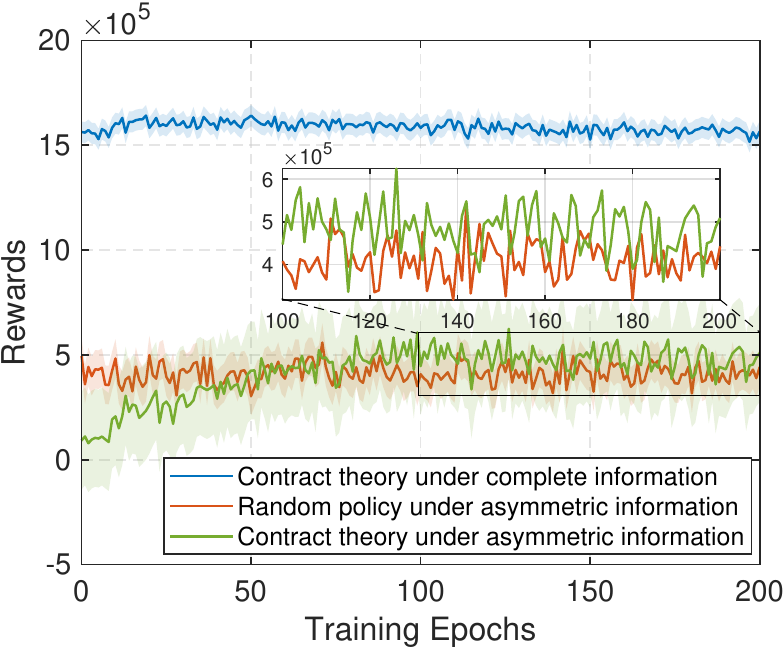}  
    \caption{Test reward comparison of the proposed scheme with the random scheme under asymmetric information and the contract theory scheme under complete information.}
    \label{Scheme_compare}
\end{figure}
\begin{figure}[!t]
    \centering  
    \includegraphics[width=0.46\textwidth]{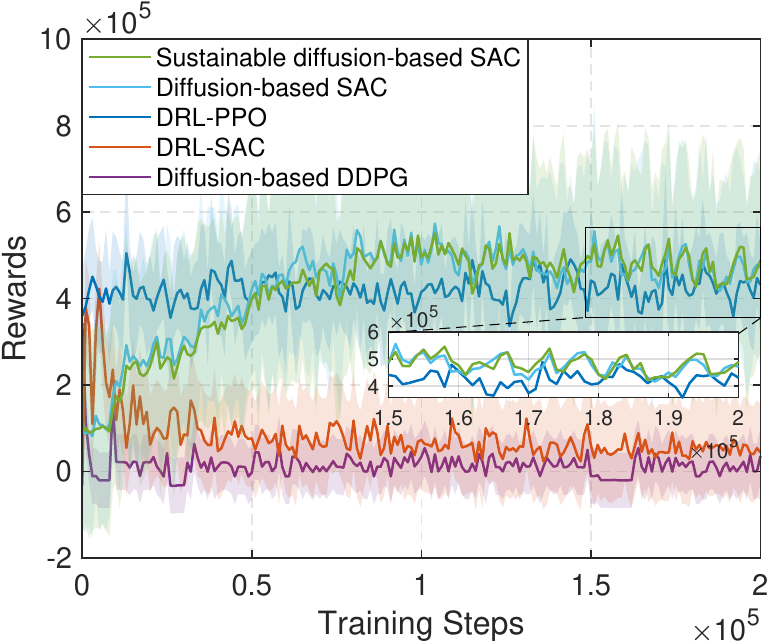}  
    \caption{Performance comparison of the proposed algorithm with several DRL algorithms in optimal contract design. For the parameter settings of the proposed algorithm, we set the pruning rate to $10\%$, the diffusion step to $6$, the learning rate of actor networks to $2\times10^{-7}$, and the learning rate of critic networks to $2\times10^{-6}$.}
    \label{Algorithm_compare}
\end{figure}
\begin{figure}[!t]
    \centering  
    \includegraphics[width=0.46\textwidth]{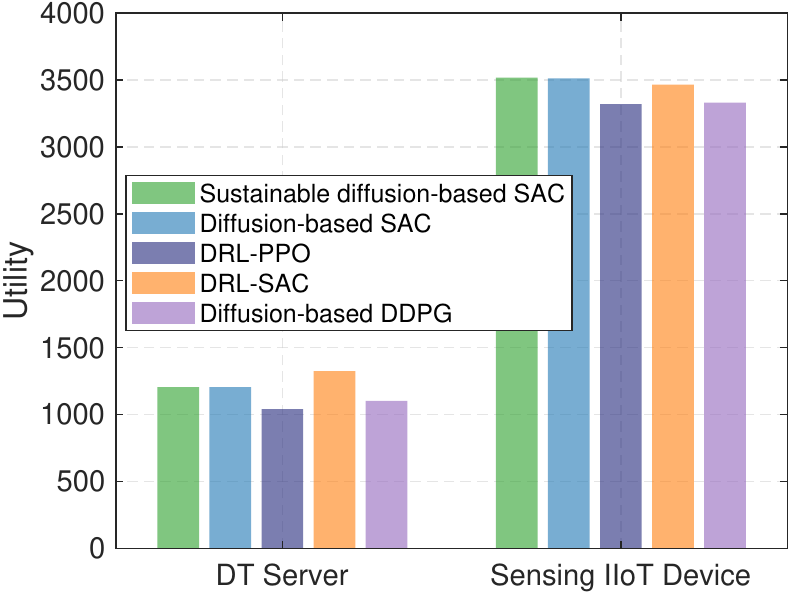} 
    \caption{The utility of the DT server and the average utility of IIoT devices under different algorithms.}
    \label{Utility}
\end{figure}

\section{Numerical Results}\label{Results}
In this section, we evaluate the performance of the proposed scheme and algorithm. We consider $M = 10$ IIoT devices and divide them into $K = 2$ types\cite{10638123, 10707303}. 

For the settings of experimental parameters, $\psi_1$ and $\psi_2$ are randomly sampled within $[50, 100]$ and $[200, 250]$, respectively. $q_1$ and $q_2$ are randomly generated following the Dirichlet distribution\cite{wen2024generativeIoT}. Considering the dynamic environment of ICPSs, the unit cost $c$ is randomly sampled within $[25,35]$, and the unit revenue $\vartheta$ is randomly sampled within $[10,15]$. In addition, the pre-defined weight parameter $\rho$ is set to $0.6$, the additional cost $c_0$ is set to $0.01$, and the pre-defined constant $\beta$ is set to $0.5$. Note that our experiments are conducted using PyTorch with CUDA 12.0 on NVIDIA GeForce RTX 3080 Laptop GPU.

Figure \ref{Scheme_compare} presents the test reward comparison of the proposed scheme under different schemes and different scenarios. Specifically, we compare the performance of the proposed scheme under asymmetric information and complete information. The contract theory under complete information does not consider IC constraints\cite{kang2023blockchain}. Although the performance of this scheme is better than that of the proposed scheme, it is not practical since the environment of complete information is not feasible. Then, we compare the performance of the proposed scheme with the random scheme in which the DT server randomly designs contracts. We can observe that the test reward of the proposed scheme is higher than that of the random scheme. The reason is that the proposed scheme can motivate IIoT devices to select suitable contract items according to their types so that the agent (i.e., the DT server) can obtain more rewards $R(\bm{e}_z,\varOmega_z)$, which indicates that the proposed scheme can effectively mitigate the effect of information asymmetry by utilizing contract theory. Overall, the proposed scheme is effective and reliable.

Figure \ref{Algorithm_compare} shows the performance of the proposed algorithm and other DRL algorithms in optimal contract design. We can observe that although the proposed algorithm does not converge quickly due to the influence of denoising, it can stabilize the highest final reward compared with other DRL algorithms. The reason is that our algorithm optimizes a stochastic policy in an entropy-augmented reward framework, encouraging exploration and robustness\cite{haarnoja2018soft}. Moreover, by leveraging diffusion models, our algorithm can generate samples with higher quality by multiple fine-tuning\cite{du2024diffusion}, enhancing the sampling accuracy and reducing the effect of uncertainty and noise from the environment. Besides, by pruning unimportant neurons and connected weights of actor networks, the performance of the diffusion-based SAC algorithm can be improved. The reason is that the pruning technique can reduce the complexity of GDM networks and improve model generalization to unseen states.

Figure \ref{Utility} illustrates the utility of the DT server and the average utility of IIoT devices under our algorithm and other DRL algorithms, where the average utility of IIoT devices is calculated based on a weighted average of device types. Due to the dynamic environment of ICPSs, we consider that the unit revenue of the DT server satisfaction is dynamically changing, and it is reasonable that the proposed algorithm achieves a higher utility of the DT server rather than the highest. However, the proposed algorithm can achieve the highest average utility of IIoT devices, indicating that the proposed algorithm can generate more reasonable contract items that better incentivize IIoT devices to contribute data for DT construction. In summary, based on the analyses of Fig. \ref{Algorithm_compare} and Fig. \ref{Utility}, the proposed algorithm can design feasible contracts to achieve the highest utility of IIoT devices under asymmetric information.

\begin{figure}[t]
    \centering  
    \includegraphics[width=0.46\textwidth]{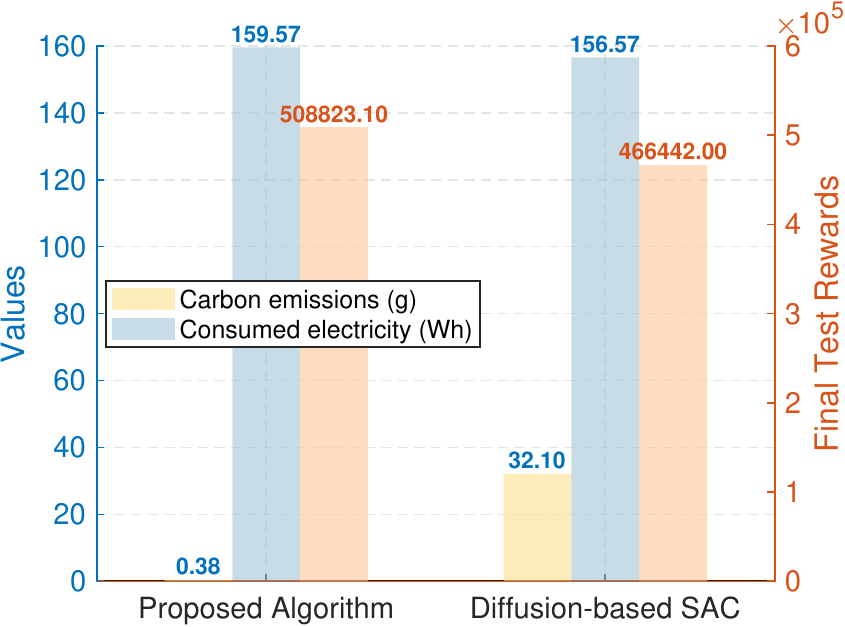}   
    \caption{Comparison of the environmental impacts of the proposed algorithm with the diffusion-based SAC algorithm.}
    \label{Carbon}
\end{figure}
\begin{figure}[t]
\centering
\subfigure[Diffusion schedules.]{
\centering
\includegraphics[width=0.46\textwidth]{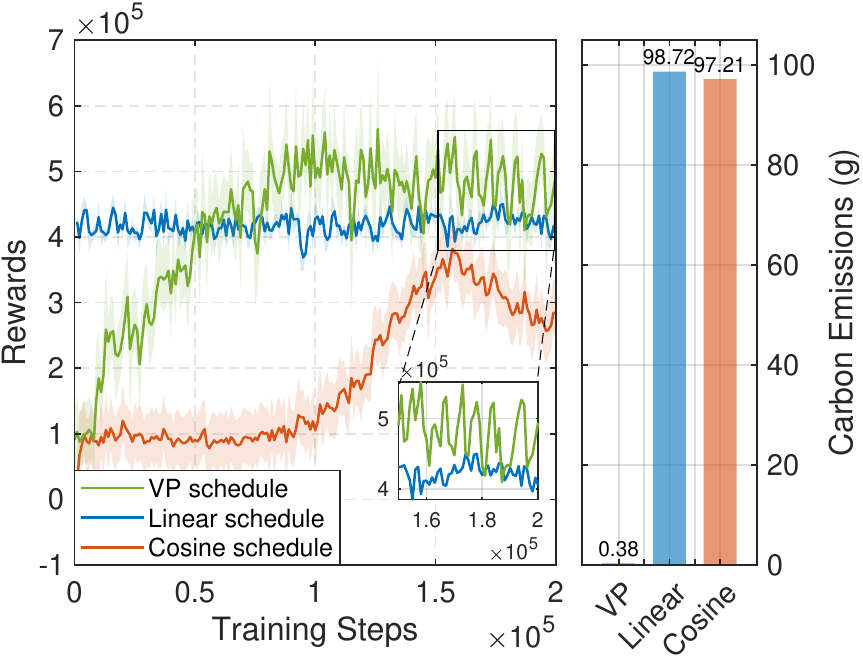}
\label{Diffusion_schedule}
}
\subfigure[Pruning rates.]{
\centering
\includegraphics[width=0.46\textwidth]{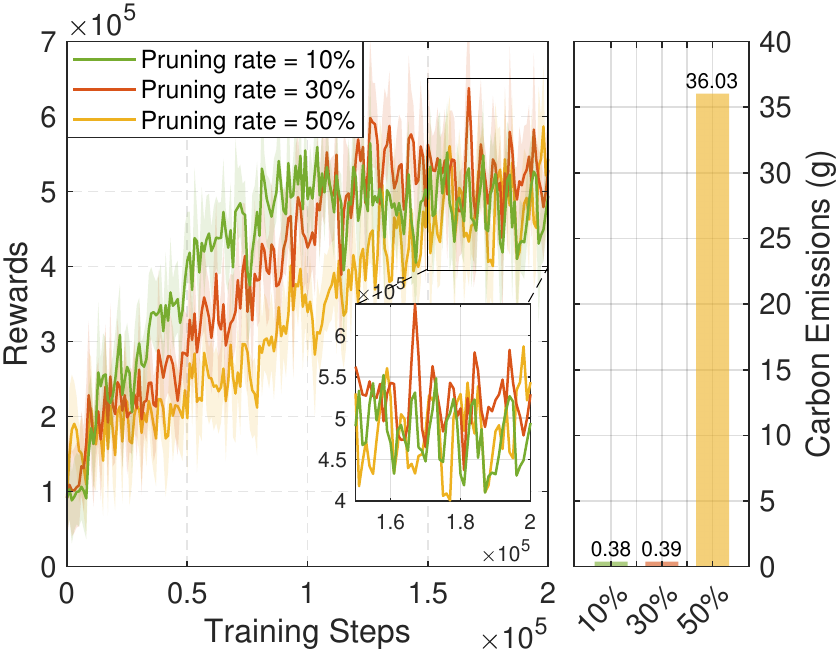}
\label{Pruning_rate}
}
\caption{Performance analyses of the proposed algorithm, where we evaluate the impacts of different noise schedule strategies and pruning rates on the performance of the algorithm.}
\label{Performance_analysis}
\end{figure}

Figure \ref{Carbon} evaluates the environmental impact of the proposed algorithm. We use a Python package named CodeCarbon\footnote{\url{https://github.com/mlco2/codecarbon}} to estimate the carbon emissions and electricity consumption of these algorithms in achieving optimal contract design. We set the pruning rate to $10\%$, which indicates that $10\%$ of channels are pruned in the whole network. Compared with the diffusion-based SAC algorithm, which also performs well in optimal contract design, we can observe that the proposed algorithm not only has better performance, i.e., a higher final test reward, but also produces lower carbon emissions of about $0.38\:\rm{g}$ during model training for optimal contract design. The reason is that pruning techniques can remove excess neurons that are useless to the performance of actor networks\cite{su2024compressing}, which is beneficial for decreasing model sizes and the need for multiple iterations, thus reducing carbon emissions. 

Figure \ref{Performance_analysis} presents the performance analyses of the proposed sustainable diffusion-based SAC algorithm. First, we evaluate the impact of different noise schedule strategies on the performance of the proposed algorithm, including Variance Proportional (VP), linear, and cosine noise schedule strategies, where the pruning rate is set to $10\%$. In diffusion models, the noise schedule strategy determines the amount of noise added to input data at each timestep during the forward diffusion process\cite{pmlr-v139-nichol21a}, which affects the quality of generated samples in the reverse diffusion process. As illustrated in Fig. \ref{Diffusion_schedule}, we can observe that among the three noise schedule strategies, the proposed algorithm under the VP noise schedule strategy can obtain the highest test reward and produce the lowest carbon emissions, which highlights the superior performance of the VP noise schedule strategy in applying the sustainable diffusion-based SAC algorithm for optimal contract design. Then, we evaluate the impact of different pruning rates on the performance of the proposed algorithm under the VP noise schedule strategy, i.e., $10\%$, $30\%$, and $50\%$. As shown in Fig. \ref{Pruning_rate}, we can observe that higher pruning rates result in slower convergence of the proposed algorithm but higher final test rewards. The reason is that higher pruning rates may eliminate critical connections that contribute to the learning ability of actor networks, resulting in slow convergence. Despite the slower convergence, the pruned actor networks often exhibit better generalization by learning more essential features instead of relying on redundant parameters. In addition, when the pruning rate is set to $50\%$, the carbon emissions generated by model training are the highest, reaching $36.03\:\rm{g}$, which indicates that there is an optimal pruning rate to balance the performance and the sustainability of the proposed algorithm. In summary, the above numerical results demonstrate that the proposed algorithm is sustainable and effective.

\section{Conclusion}\label{Conclusion}
In this paper, we have studied how GenAI empowers DTs in ICPSs. Specifically, we have designed a GenAI-driven DT architecture in ICPSs, which systematically studies how GenAI drives the DT construction pipeline, including real-time physical data collection, communications for DTs, DT modeling and maintenance, and DT decision-making. To motivate IIoT devices to contribute sensing data for GenAI-empowered DT construction, we have proposed a contract theory model under information asymmetry. Furthermore, we have developed a sustainable diffusion-based SAC algorithm to generate the optimal feasible contract, which utilizes dynamic structured pruning techniques to sparsify actor networks of GDMs, allowing efficient implementation of the proposed algorithm in ICPSs. Finally, the numerical results demonstrate that the proposed algorithm outperforms DRL algorithms that are commonly used for optimal contract design. In particular, when the pruning rate is set to $10\%$, compared with the diffusion-based SAC algorithm, the proposed algorithm can reduce carbon emissions by $99\%$ during model training while ensuring high test rewards. For future work, we will propose a more general and comprehensive architecture of GenAI-empowered DT in ICPSs. In addition, we will focus on the incentive mechanism design among multiple IIoT devices and DT servers and explore combining state-of-the-art techniques with DRL algorithms for optimal contract design.

\bibliographystyle{Bibliography/IEEEtranTIE}
\bibliography{ref}

\begin{IEEEbiography}
[{\includegraphics[width=1in,height=1.25in,clip,keepaspectratio]{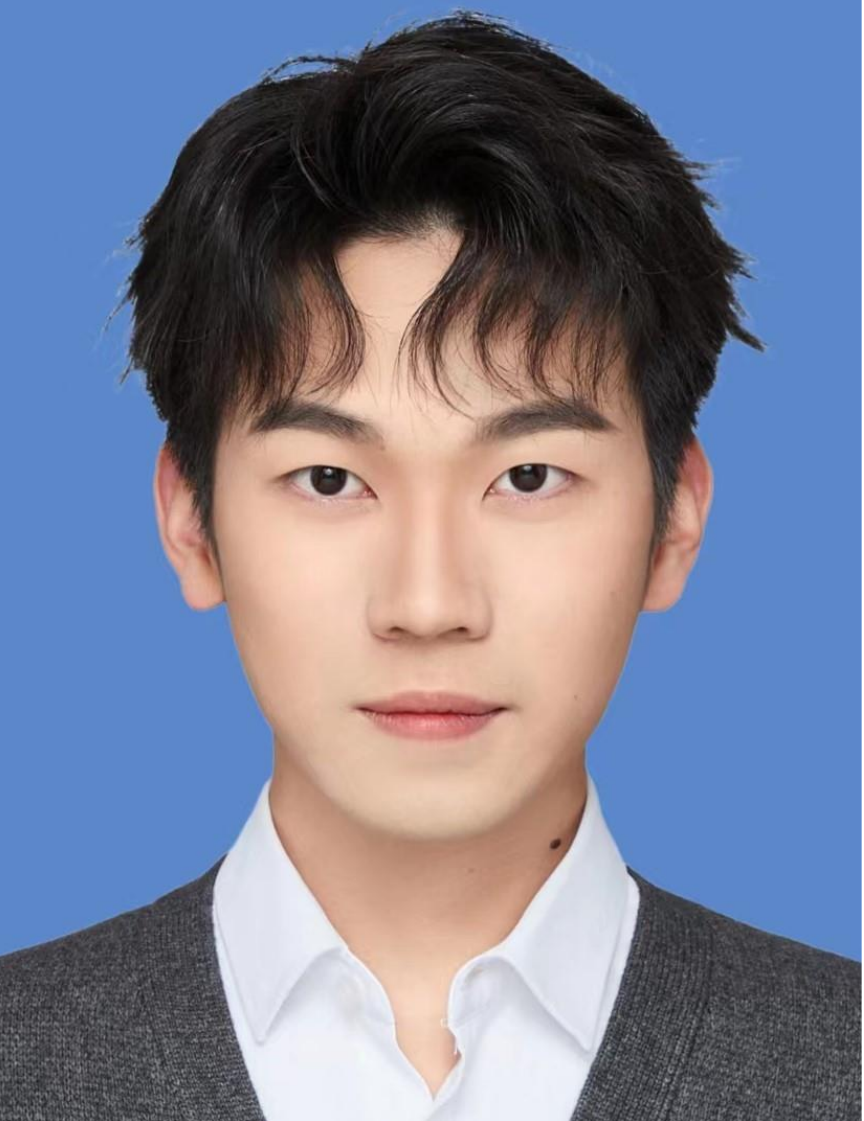}}] 
{Jinbo Wen} received the B.Eng. degree from Guangdong University of Technology, China, in 2023. He is currently pursuing an M.S. degree with the College of Computer Science and Technology, Nanjing University of Aeronautics and Astronautics, China. His research interests include generative AI, edge intelligence, and metaverse.
\end{IEEEbiography}

\begin{IEEEbiography}[{\includegraphics[width=1in,height=1.25in,clip,keepaspectratio]{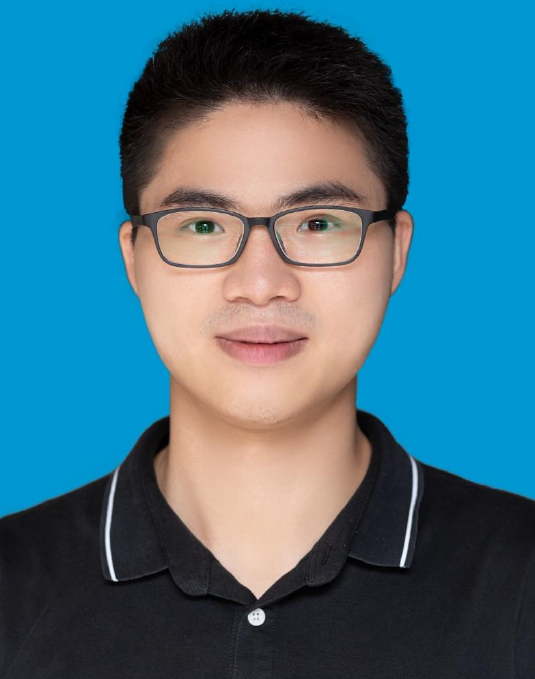}}]{Jiawen Kang} received the Ph.D. degree from Guangdong University of Technology, China, in 2018. He has been a postdoc at Nanyang Technological University, Singapore from 2018 to 2021. He is currently a full professor at Guangdong University of Technology, China. His research interests mainly focus on blockchain, security, and privacy protection in wireless communications and networking.
\end{IEEEbiography}

\begin{IEEEbiography}
[{\includegraphics[width=1in,height=1.25in,clip,keepaspectratio]{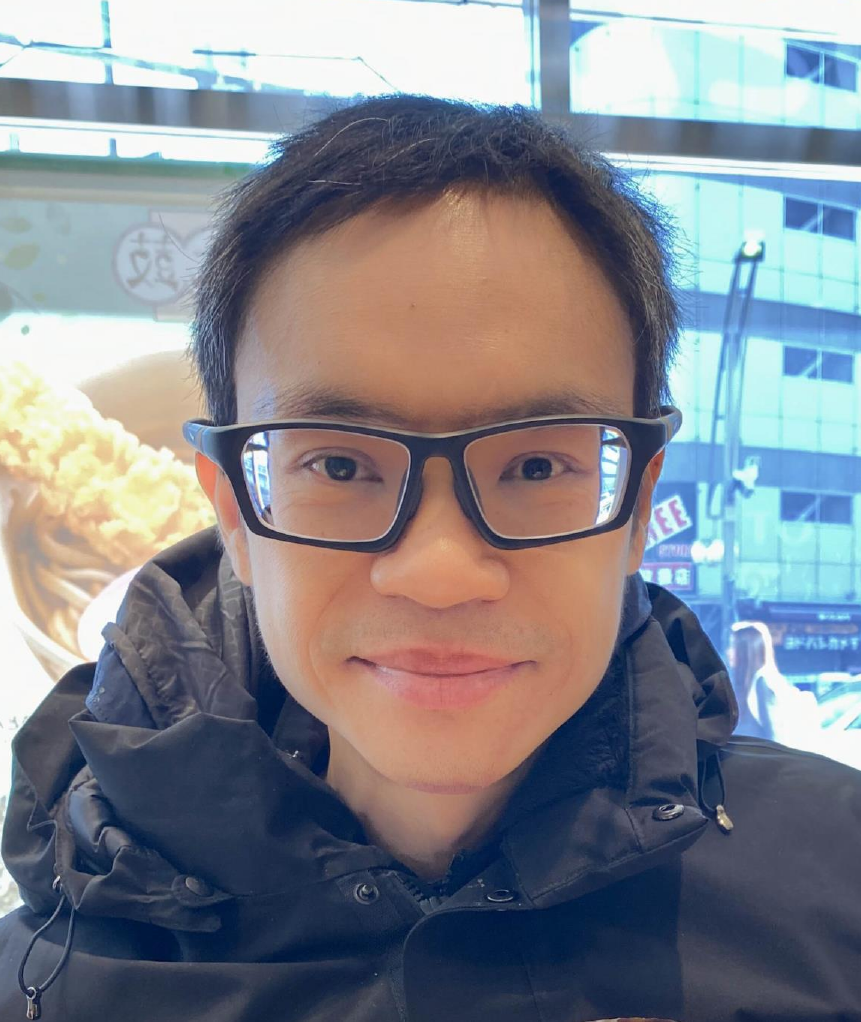}}] 
{Dusit Niyato} (Fellow, IEEE) is a professor in the College of Computing and Data Science, at Nanyang Technological University, Singapore. He received the B.Eng. degree from King Mongkuts Institute of Technology Ladkrabang (KMITL), Thailand in 1999 and the Ph.D. in Electrical and Computer Engineering from the University of Manitoba, Canada in 2008. His research interests are in the areas of sustainability, edge intelligence, decentralized machine learning, and incentive mechanism design.
\end{IEEEbiography}

\begin{IEEEbiography}
[{\includegraphics[width=1in,height=1.25in,clip,keepaspectratio]{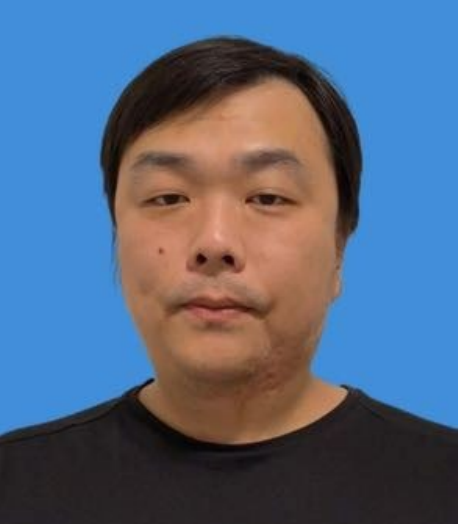}}] 
{Yang Zhang} is currently an associate professor in the College of Computer Science and Technology, Nanjing University of Aeronautics and Astronautics, Nanjing, China. He received B. Eng. and M. Eng. from Beihang University in 2008 and 2011, respectively. He obtained the Ph.D. degree in Computer Engineering from Nanyang Technological University, Singapore, in 2015. He is an editor of the IEEE Transactions on Machine Learning in Communications and Networking. His current research topic is edge computing and multi-agent unmanned systems.
\end{IEEEbiography}

\begin{IEEEbiography}
[{\includegraphics[width=1in,height=1.25in,clip,keepaspectratio]{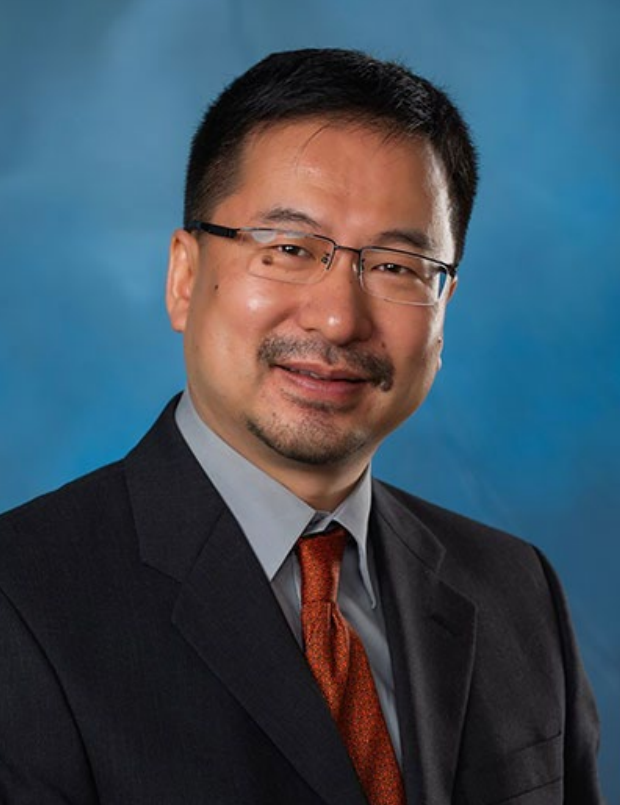}}] 
{Shiwen Mao} (Fellow, IEEE) received his Ph.D. in electrical and computer engineering from Polytechnic University, Brooklyn, NY. He is a Professor and Earle C. Williams Eminent Scholar and Director of the Wireless Engineering Research and Education Center at Auburn University. His research interests include wireless networks and multimedia communications. He is a Distinguished Lecturer of the IEEE Communications Society and IEEE Council of RFID. He is the editor-in-chief of IEEE Transactions on Cognitive Communications and Networking, a Member-at-Large of IEEE Communications Society Board of
Governors, and Vice President of Technical Activities of IEEE Council on Radio Frequency Identification (CRFID). He was the General Chair of IEEE INFOCOM 2022, a TPC Chair of IEEE INFOCOM 2018, and a TPC Vice-Chair of IEEE GLOBECOM 2022. He received the IEEE ComSoc MMTC Outstanding Researcher Award in 2023, the SEC 2023 Faculty Achievement Award for Auburn, the IEEE ComSoc TC-CSR Distinguished Technical Achievement Award in 2019, the Auburn University Creative Research \& Scholarship Award in 2018, and the NSF CAREER Award in 2010, as well as several IEEE service awards. He is a co-recipient of the 2022 Best Journal Paper Award of IEEE ComSoc eHealth Technical Committee, the 2021 Best Paper Award of Elsevier/KeAi Digital Communications and Networks Journal, the 2021 IEEE Internet of Things Journal Best Paper Award, the 2021 IEEE Communications Society Outstanding Paper Award, the IEEE Vehicular Technology Society 2020 Jack Neubauer Memorial Award, the 2018 Best Journal Paper Award and the 2017 Best Conference Paper Award from IEEE ComSoc MMTC, and the 2004 IEEE Communications Society Leonard G. Abraham Prize in the Field of Communications Systems.
\end{IEEEbiography}

\end{document}